\documentclass[sigconf]{acmart}

\AtBeginDocument{%
  }

\copyrightyear{2026}
\acmYear{2026}
\setcopyright{cc}
\setcctype{by}
\acmConference[ICAIF '26]{7th ACM International Conference on AI in Finance}{November 14--17, 2026}{Milan, AA, Italy}
\acmBooktitle{7th ACM International Conference on AI in Finance (ICAIF '26), November 14--17, 2026, Milan, AA, Italy}
\acmDOI{10.1145/3858063.3858790}
\acmISBN{979-8-4007-3087-0/2026/11}

\usepackage{algorithm}
\usepackage{multirow}
\usepackage{caption}
\usepackage[printonlyused]{acronym}
\usepackage{tikz}
\usepackage{animate}
\usepackage{booktabs}
\usepackage{algorithm}
\usepackage{algpseudocode}

\usetikzlibrary{arrows.meta}
\usetikzlibrary{positioning}
\usetikzlibrary{fit}
\usetikzlibrary{calc}
\usetikzlibrary{shapes.geometric}

\definecolor{customred}{RGB}{165,70,70}
\definecolor{customredfill}{RGB}{248,235,235}

\definecolor{customblue}{RGB}{65,105,160}
\definecolor{custombluefill}{RGB}{232,239,249}

\definecolor{customgreen}{RGB}{65,130,105}
\definecolor{customgreenfill}{RGB}{232,246,240}

\definecolor{customorange}{RGB}{185,115,45}
\definecolor{customorangefill}{RGB}{250,240,226}

\definecolor{customdarkblue}{RGB}{70,95,150}
\definecolor{customdarkbluefill}{RGB}{235,240,250}

\definecolor{custompink}{RGB}{175,85,100}
\definecolor{custompinkfill}{RGB}{250,236,240}

\definecolor{custompurple}{RGB}{95,80,145}
\definecolor{custompurplefill}{RGB}{240,236,248}

\definecolor{customgray}{RGB}{95,94,90}
\definecolor{customgrayfill}{RGB}{242,242,240}

\DeclareMathOperator{\diag}{diag}

\DeclareMathOperator{\tr}{tr}

\DeclareMathOperator{\Cov}{Cov}

\DeclareMathOperator{\offdiag}{offdiag}

\newcommand{\SampleCovar}{\hat{\boldsymbol{\Sigma}}_S}
\newcommand{\FactorCovar}{\hat{\boldsymbol{\Sigma}}_B}
\newcommand{\RepCovar}{\hat{\boldsymbol{\Sigma}}_R}

\newcommand{\frameworkShort}{MINGLE}
\newcommand{\frameworkFull}{%
  \underline{\textbf{M}}utually--%
  \underline{\textbf{IN}}formed %
  \underline{\textbf{G}}raph--%
  \underline{\textbf{L}}ocality %
  and %
  \underline{\textbf{E}}xposures%
}
\newcommand{\frameworkLong}{\frameworkFull{} framework (\frameworkShort{})}

\algnewcommand{\Input}{\item[\textbf{Input:}]}

\newcounter{remark}
\newcommand{\remark}[1]{%
  \refstepcounter{remark}%
  \noindent \textbf{Remark \theremark.} #1%
}

\begin{document}
\acrodef{ADMM}{Alternating Direction Method of Multipliers}
\acrodef{APT}{Arbitrage Pricing Theory}
\acrodef{BIC}{Bayesian Information Criterion}
\acrodef{CAGR}{Compound Annual Growth Rate}
\acrodef{CutN}{Cardinality-normalised Graph Cut}
\acrodef{CutV}{CutV}
\acrodef{ENB}{Effective Number of Bets}
\acrodef{EWMA}{Exponentially-Weighted Moving Average}
\acrodef{FYP}{Final Year Project}
\acrodef{GICS}{Global Industry Classification Standard}
\acrodef{GMM}{Gaussian Mixture Model}
\acrodef{GMRF}{Gaussian Markov Random Field}
\acrodef{HRP}{Hierarchical Risk Parity}
\acrodef{MPT}{Modern Portfolio Theory}
\acrodef{MV}{Minimum Variance}
\acrodef{MVO}{Mean-Variance Optimisation}
\acrodef{NMI}{Normalised Mutual Information}
\acrodef{PCA}{Principal Component Analysis}
\acrodef{PR}{Participation Ratio}
\acrodef{RBF}{Radial Basis Function}
\acrodef{RMT}{Random Matrix Theory}
\acrodef{SNR}{Signal to Noise Ratio}
\acrodef{SVD}{Singular Value Decomposition}
\acrodef{VIX}{Volatility Index}

\title{Beyond Co-Movement: Locality by Exposures Enables a Joint Factor--Graph Framework for Portfolio Diversification}

%% Authors list
\author{Sara Chehab}
\authornote{These authors contributed equally to this work.}
\orcid{0009-0001-9463-8376}
\email{sara.chehab22@imperial.ac.uk}
\affiliation{%
  \institution{Imperial College London}
  \city{London}
  \country{UK}
}

\author{Giorgos Iacovides}
\email{giorgos.iacovides20@imperial.ac.uk}
\orcid{0009-0007-5733-8992}
\authornotemark[1]
\affiliation{%
  \institution{Imperial College London}
  \city{London}
  \country{UK}
}

\author{Parisa Yazdanparast}
\email{parisa.yazdan@gmail.com}
\orcid{0009-0002-9001-0335}
\affiliation{%
  \institution{Beaufort Bridges}
  \city{London}
  \country{UK}
}

\author{Danilo Mandic}
\orcid{0000-0001-8432-3963}
\email{d.mandic@imperial.ac.uk}
\affiliation{%
  \institution{Imperial College London}
  \city{London}
  \country{UK}
}

% \renewcommand{\shortauthors}{Chehab et al}

\begin{abstract}
  Current portfolio construction methods are either agnostic to the effects of idiosyncratic shocks (standard factor models) or to the latent data structure driving systematic returns (recent graph-based approaches). This presents an opportunity to combine the complementary market aspects captured by the factor and graph domains, allowing asset allocations to operate directly on the underlying market structure, rather than on its observed co-movement or its finite-sample artefacts.
In this work, we introduce the \frameworkLong, which mutually regularises the factor and graph domains by redefining graph locality through systematic factor exposure profiles, rather than via observed co-movements. This is formalised through a unified \acf{ADMM} framework that jointly learns a latent factor representation and its induced graph topology directly from market returns. The resulting exposure-similarity graph aligns more closely with established economic sectors than conventional correlation-based graphs.
Portfolios constructed from this representation are shown to consistently outperform their correlation-based counterparts across a range of volatility regimes and transaction cost levels.
For rigour, paired statistical testing confirms that these gains stem from the reconciliation of the graph and factor domains.
\end{abstract}



% \received{2 August 2026}
% \received[revised]{2 August 2026}
% \received[accepted]{2 August 2026}

\maketitle

\section{Introduction}
Portfolio diversification aims to maximise profit, and as such rests upon an accurate characterisation of inter-asset relationships, inferred from historical returns. In this setting, return co-movements serve as a proxy for the underlying market structure, since assets with similar factor exposures are expected to exhibit correlated returns under an \ac{APT} based factor decomposition.

The characterisation of asset co-movement is traditionally performed through standard similarity metrics, such as the sample covariance. However, such second-order similarity metrics tend to be unreliable under scenarios with pronounced uncertainty or shifts in dynamical regimes.
Paradoxically, the covariance metric becomes a particularly unreliable proxy for market structures during crisis periods, when diversification is most needed. This phenomenon undermines canonical \ac{MVO} approaches~\cite{markowitz_portfolio_1952}, and is known as the \emph{Markowitz curse}~\cite{lopez_de_prado_building_2016}.

To mitigate the errors in covariance estimation, various inductive biases have been proposed to regularise inter-asset relationships, including additional portfolio constraints~\cite{jagannathan_risk_2003,demiguel_generalized_2009}, Bayesian priors~\cite{black_global_1992,jorion_bayes-stein_1986}, and graph-based representations~\cite{mantegna_hierarchical_1999,peralta_network_2016}. Existing graph-based approaches, such as spectral clustering \cite{dees_portfolio_2019,dionysiou_adaptive_2026}, explore risk diffusion among the local neighbourhoods of co-moving assets. However, graph topologies inferred from the empirical covariance tend to conflate systematic and idiosyncratic sources of risk. Consequently, irrespective of estimation quality, covariance-based graphs cannot attribute an observed co-movement to a macroeconomic market structure or a localised shock. On the other hand, standard factor-based models~\cite{bodie_arbitrage_2023} decompose returns into separable and interpretable risk factors. However, while such models are successful in isolating broad risks by design, they cannot capture the shared, peer-to-peer responsiveness across the asset universe that graph-based methods natively encode.

Despite their complementary nature, these two views remain largely siloed. It is then natural to ask ourselves:
\begin{itemize}
    \item \textit{Can we reconcile both systematic risk drivers and co-responses to perturbations to obtain a richer market representation and provide a stronger basis for portfolio diversification?}
\end{itemize}
We argue that portfolio diversification is best achieved by grouping assets according to their \emph{exposure to latent sources of risk}, rather than via their observed co-movement, ensuring that the resulting neighbourhoods capture persistent economic structure instead of transient, window-specific correlations.
To this end, we introduce the \frameworkLong, which jointly learns a factor model and graph topology directly from raw returns. Specifically, we leverage the learnt exposures to equip the graph domain with a locality measure which is grounded in systematic risk rather than in raw co-movement. Conversely, we leverage the learnt graph to equip the factor domain with exposures that remain consistent across neighbourhoods of co-responsive assets. Such \emph{domain-informed} \emph{mutual consistency} yields diversification gains beyond the capability of either domain in isolation.

The main contributions of this work are therefore:
\begin{itemize}
    \item We propose the \acs{ADMM}-based \frameworkShort\ framework that jointly, rather than independently, estimates a latent factor model and its induced graph topology directly from market returns. By redefining graph locality through shared factor exposures, rather than via observed co-movements, \frameworkShort\ captures persistent economic structure. % and entirely bypasses the need for intermediate covariance estimates. % The resulting optimisation is formulated within a unified \acf{ADMM} scheme.
    \item We demonstrate that the covariance structure induced by the \frameworkShort\ representation is substantially better conditioned than the standard sample covariance. This provides a significantly more stable and discriminative mathematical foundation for downstream portfolio optimisation.
    \item We show that portfolios constructed upon the \frameworkShort\ exposure similarity graphs consistently outperform their correlation-based baselines across different market regimes. We also confirm that the performance gains stem fundamentally from the improved market representation, rather than the choice of the allocator.
\end{itemize}

\section{Related Works} \label{sec:bg}
In high-dimensional financial settings with $p$ assets, the available observations,~$T$, are usually statistically insufficient to reliably estimate the sample covariance of pairwise co-movements,~\mbox{$\SampleCovar \in \mathbb{R}^{p \times p}$}.
Indeed, while strong trends associated with large eigenvalues naturally emerge from the data, smaller trends can be conflated with artefacts. These small eigenvalues, which are statistically indistinguishable from noise, lie in the so-called Marchenko--Pastur bulk~\cite{marcenko_distribution_1967,plerou_random_2002}, and induce the ill-conditioning of~$\SampleCovar$.

This ill-conditioning worsens during crisis regimes, when cross-asset correlations spike uniformly~\cite{longin_extreme_2001}, and a greater share of the spectrum collapses into the Marchenko--Pastur bulk. As a result, allocations relying on the inversion of~$\SampleCovar$, such as \ac{MVO}, are increasingly ill-posed exactly when diversification is most needed. This paradox is colloquially known as the \emph{Markowitz curse}~\cite{lopez_de_prado_building_2016}.

To mitigate the ill-conditioning of~$\SampleCovar$, structural priors may be imposed, deliberately introducing inductive biases to achieve a substantial reduction in estimation variance. These inductive biases, rather than arbitrary regularisations, encode domain knowledge and priors about the structure and dynamics of financial markets.

\subsection{Graph Topologies}

Financial assets are known to exhibit \emph{local} structural similarity: small neighbourhoods of assets behave alike, while progressively larger, looser neighbourhoods capture increasingly systematic, market wide trends. This multi-scale view of the market is naturally encoded by graph topologies, where the strength of the relationship between two assets defines their proximity.

Capital allocation can employ pre-existing locality measures which indicate how localised perturbations propagate within neighbourhoods of increasing sizes. Among these, spectral cut methods, such as \acs{CutV}~\cite{dees_portfolio_2019}, exploit the graph Laplacian eigenspectrum to identify information bottlenecks and recursively bipartition the graph. While the theoretically optimal clustering relies on the Fiedler vector~\cite{fiedler_algebraic_1973}, its reliability degrades when
the eigengaps around the corresponding eigenvalue are small~\cite{davis_rotation_1970}. To mitigate the presence of artefacts, recent works have proposed selecting the partition eigenvector through an information-theoretic criterion~\cite{dionysiou_adaptive_2026}. 

However, a graph is only as informative as the notion of locality used for its construction: locality defines the topology of a graph, and any inadequacy in this underlying locality measure is inevitably inherited by methods operating on the graph.
More specifically, spectral cuts, which tend to employ correlation-based graphs, conflate direct pairwise relationships with transitive, wider topological dependencies.
If locality is central to the quality of the resulting representation and its downstream application, it can be treated as a \emph{modelling decision}, whereby a \emph{smoothness prior}~\cite{dong_learning_2016} may be used to encourage assets with similar representations to lie close together and discourage links between assets with dissimilar representations,~$\mathbf{y}_i$ and $\mathbf{y}_j$. This rationale can be formalised through the minimisation of the Dirichlet energy~\cite{dong_laplacian_2015}, in the form
\begin{equation}
    \tr \left( \mathbf{Y}^\top \mathbf{L} \mathbf{Y} \right) = \frac{1}{2} \sum_{i=1}^p \sum_{j=1}^p \; w_{ij} \Vert \mathbf{y}_i - \mathbf{y}_j \Vert_2^2
\end{equation}
where $\mathbf{Y} \in \mathbb{R}^{p\times T}$ denotes the vertex representation, $\mathbf{W} \in \mathbb{R}^{p\times p}$ the graph weight matrix, and $\mathbf{L} = \diag \left( \mathbf{W1}\right) - \mathbf{W}$ the graph Laplacian, with $\mathbf{W1}$ as the vector of node degrees. The vertex representation and graph topology can be jointly learnt by imposing locality through the optimisation problem 
\begin{equation}
    \min_{\mathbf{Y}, \; \mathbf{L} \in \mathcal{L}} \; \Vert \mathbf{X} - \mathbf{Y} \Vert_F^2 + \lambda \; \tr \left( \mathbf{Y}^\top \mathbf{L} \mathbf{Y} \right) \label{eq:bg_graph_learning}
\end{equation}
where $\lambda$ controls the strength of the smoothness prior imposed on the learnt vertex representation, $\mathbf{Y}$, over the learnt graph topology,~$\mathbf{W}$, while $\mathcal{L}$ denotes the admissible set of graph Laplacians.
The regularised optimisation task causes the graph to adapt its neighbourhood structure according to $\mathbf{Y}$. Simultaneously, $\mathbf{Y}$ is  encouraged to remain close to the observed returns, $\mathbf{X}$, and be smooth over the graph, but remains otherwise unconstrained. Imposing further structure onto $\mathbf{Y}$ requires additional market priors.

% METHODOLOGY FIGURE + VISUAL ABSTRACT MOVED TO BACKGROUND SECTION, FOR IT TO APPEAR ON TOP OF THE PAGE
\begin{figure*}[t]
    \centering
    \input{equations}
    \Description{}
    \caption[Overview of the proposed factor--graph model]{The proposed \frameworkShort\ framework jointly learns a factor model, $\hat{\mathbf{X}} \approx \mathbf{B}\mathbf{F}$, and a graph topology, $\mathbf{W}$, from the raw asset returns, $\mathbf{X} \in \mathbb{R}^{p\times T}$. In the factor model, $\mathbf{B} \in \mathbb{R}^{p \times k}$ encodes the exposure profiles to $k$ latent risk drivers, whereby the height of the bar reflects the exposure magnitude and its colour reflects the corresponding factor, and $\mathbf{F} \in \mathbb{R}^{k \times T}$ captures the corresponding factor dynamics. These exposure profiles are defined to be smooth on the learnt graph, $\mathbf{W} \in \mathbb{R}^{p \times p}$, such that the strength of the relationship between two assets reflects their exposure similarity, rather than their observed co-movement. The factor and graph representations encode complementary views of the market and are jointly learnt, with neither block estimated in isolation, to capture both the causes of systematic co-movements and the co-responsive behaviour to common risk drivers.}
    \label{fig:method_diagram}
\end{figure*}

\subsection{Factor Models}

Asset returns are commonly modelled as arising from a small number of common latent risk drivers. The \acf{APT}~\cite{bodie_arbitrage_2023} formalises this intuition by modelling the return of an asset $i$ across time as arising from contributions of $k$ factors, in the form
\begin{equation}
    \mathbf{X} = \mathbf{B} \mathbf{F} + \boldsymbol{\varepsilon}, \label{eq:bg_apt_latent_matrix}
\end{equation}
where $\boldsymbol{\varepsilon}$ designates the idiosyncratic components, $\mathbf{F}$ the factor dynamics, and $\mathbf{B}$ the asset exposures to these factors. When $\mathbf{B}$ and $\mathbf{F}$ are estimated using the first $k$ principal components of the covariance of historical returns, the factor model induces a covariance
\begin{equation}
    \FactorCovar = \mathbf{B} \hat{\boldsymbol{\Sigma}}_F \mathbf{B}^\top + \boldsymbol{\Psi}, \label{eq:bg_cov_apt}
\end{equation}
where $\hat{\boldsymbol{\Sigma}}_F = \Cov(\mathbf{f}_t)$ and $\boldsymbol\Psi = \diag\left( \psi_1, \, \; \ldots, \; \psi_p  \right)$ captures the idiosyncratic variance of each asset. This makes the resulting covariance matrix, $\FactorCovar$, more resilient to artefacts than $\SampleCovar$, as its systematic component $\mathbf{B} \boldsymbol\Sigma_F \mathbf{B}^\top$ is low-rank: it only retains the $k$ largest eigenvalues and discards the smaller ones attributable to insignificant information or noise.

While factor models address the statistical estimation problem of~$\SampleCovar$, they expose a deeper structural limitation of the covariance: the systematic component, encoding stable inter-asset relationships, and the idiosyncratic noise, encoding transient asset-specific shocks, are observed only through their aggregate effect. Locality derived from correlation therefore conflates genuine shared exposure with idiosyncratic components, irrespective of estimation quality.

\subsection{Joint Factor-Graph Representations}

Traditional market representations have encoded one of two complementary views: factor models capture the sources that drive returns but remain agnostic to inter-asset relationships, while graph representations capture these inter-asset relationships but remain agnostic to systematic components. Joint approaches have attempted to reconcile both market views \cite{martin_factor-driven_2025}, though their sequential, decoupled learning paradigms have kept the domains from informing and regularising each other. 
Indeed, in a sequential scheme, the factor domain is estimated first and is held fixed during graph estimation. As such, the graph domain cannot correct the bias induced by the factor domain. It is therefore only a coupled estimation that allows the graph topology and factor domain to mutually regularise one another and become \emph{mutually aware}.
To this end, in the present paper, we propose \frameworkShort, a unified \ac{ADMM} framework that enforces \emph{mutual consistency} across the factor and graph domains so that each corrects for the inductive bias of the other.

\section{Methodology} \label{sec:method}
% METHODOLOGY FIGURE + VISUAL ABSTRACT MOVED TO BACKGROUND SECTION, FOR IT TO APPEAR ON TOP OF THE PAGE

Assets with similar exposures to latent risk drivers are expected to respond similarly to systematic market perturbations and therefore offer limited diversification benefit to one another. For the purpose of diversification, we define graph locality in the factor-exposure domain, with asset neighbourhoods reflecting similarity in exposure profiles, rather than in-sample co-movements. This definition of locality implies a \emph{mutual consistency} of the graph and factor domains, whereby the factor model determines the exposure profiles over which locality is measured, while the graph regularises those profiles according to the neighbourhood structure it induces. This mutual consistency underpins the \frameworkShort\ framework, and is visualised in~\autoref{fig:method_diagram}.

\subsection{Unified Objective Formalisation}

For a universe with $p$ assets and $T$ observations, we ensure the factor and graph domains become mutually aware by estimating them jointly within a unified framework, rather than independently in separate stages. We formalise this factor--graph coupling by regularising the unconstrained representation of~\eqref{eq:bg_graph_learning} to be low-rank, $\mathbf{Y} \approx \mathbf{BF}$ as in~\eqref{eq:bg_apt_latent_matrix}, and by enforcing exposure smoothness on the learnt topology, $\mathbf{W}$. The resulting \frameworkShort\ objective is given by
\begin{equation}
    \begin{aligned}
        \min_{\mathbf{B},\,\mathbf{F},\,\mathbf{W}} \quad
            & \underbrace{\lVert \mathbf{X} - \mathbf{B}\mathbf{F} \rVert_{\boldsymbol{\Omega}}^2}_{\text{Factor decomposition}}
                + \underbrace{\lambda \sum_{i=1}^p \sum_{j=1}^p
                w_{ij}\,\lVert \mathbf{b}_{i} - \mathbf{b}_{j} \rVert_2^2}_{\text{Exposure--smoothness on graph}} \\[0.1em]
            & + \underbrace{\alpha \lVert \mathbf{W} \rVert_1}_\text{Graph sparsity}
                - \underbrace{\beta \; \mathbf{1}^\top \log \left(\mathbf{W}\mathbf{1} + \varepsilon \mathbf{1} \right)}_{\text{Degree barrier}} \\[0.2em]
            & + \underbrace{\delta \Vert \offdiag(\mathbf{F} \boldsymbol\Omega \mathbf{F}^\top) \Vert_F^2}_{\text{Factor decorrelation}} \\[0.3em]
        \text{s.t.} \quad
            & \underbrace{\mathbf{B}^\top \mathbf{B} = \mathbf{I},}_{\text{\acs{PCA} scale ambiguity}}\; \underbrace{ w_{ij} \geq 0 \; \forall i, j, \; w_{ii} = 0 \; \forall i, \; \mathbf{W}^\top = \mathbf{W}}_{\text{Similarity graph topological constraints}},
    \end{aligned}
    \label{eq:method_unified_obj}
\end{equation}
where $\offdiag({\mathbf{X}}) = \mathbf{X} - \diag(\mathbf{X})$; $\boldsymbol{\Omega} \in \mathbb{R}^{T \times T}$ is a diagonal matrix enforcing exponential time-decay; while $\mathbf{X} \in \mathbb{R}^{p\times T}$, $\mathbf{F} \in \mathbb{R}^{k \times T}$, \mbox{$\mathbf{B} \in \mathbb{R}^{p \times k}$}, and $\mathbf{W} \in \mathbb{R}^{p \times p}$ denote respectively the historical returns, factor dynamics, factor exposures, and graph topology. The hyperparameters and their respective roles are defined in~\autoref{tab:implementation_hyperparameters}.

The first two terms in~\eqref{eq:method_unified_obj} are the core of the proposed framework, with the reconstruction term recovering a small number, $k$, of risk drivers from the historical returns and the smoothness term coupling the resulting exposures with the graph topology. 
As neither $\mathbf{B}$ nor $\mathbf{W}$ is fixed \emph{a priori}, the graph and factor domains correct for each other's inductive biases, as elaborated in~\autoref{sec:bg}. The graph regularises the low-rank decomposition beyond variance maximisation, toward exposures that mirror co-responsive neighbourhood structures; the factor model regularises the graph domain beyond observed co-movement, toward a topology whose neighbourhoods are defined by similarity in systematic risk exposure. The smoothness term acts as the shared inductive bias and thus ensures cross-domain awareness: for a fixed $\mathbf{W}$, it is imposed on the exposure space to promote similar exposures in neighbourhoods of co-responsive assets; conversely, for a fixed $\mathbf{B}$, it is imposed on the topology to penalise edges between assets with dissimilar exposures.
The remaining terms of the unified objective regularise the factor and graph domains according to a set of economically motivated considerations and guard against degenerate solutions.
The factor-domain regularisers first encourage non-redundant systematic risk sources by penalising the off-diagonal elements of the induced factor covariance matrix,~$\mathbf{F}\boldsymbol{\Omega}\mathbf{F}^\top$.
Within the graph domain, spurious relationships attributable to artefacts are dropped through the sparsity-inducing $\ell_1$ penalty on $\mathbf{W}$.
Additionally, assets are encouraged to belong to a non-trivial neighbourhood to ensure that their risk remains diversifiable; this graph connectivity is enforced through the log-barrier preventing vanishing node degrees.
Further topological constraints enforce graph similarity properties and yield a positive semi-definite graph Laplacian, as required by the Adaptive \acs{CutV} algorithm~\cite{dionysiou_adaptive_2026}.

\subsection{Joint Learning Process}

\begin{algorithm}
    \small
    \caption{\acs{ADMM} iterations for solving the split optimisation problem of~\eqref{eq:method_obj_aux} through its augmented Lagrangian, alternating between the factor, graph, and dual updates.
    Only the exposure-profile update, for $\mathbf{B}$, requires a nested iterative solver, with its $\mathcal{O}(p^3)$ Sylvester solve dominating the per-iteration cost. The gradient step used during the $\mathbf{F}$ update costs $\mathcal{O}(pkT)$, while the remaining updates admit direct or closed-form solutions and cost at most $\mathcal{O}(p^2)$.
    }
    \label{algo:method_steps_admm}
    \begin{algorithmic}
    \Input{$\mathbf{B}^0$, $\mathbf{F}^0$, $\mathbf{W}^0$, $\mathbf{Q}^0$, $\mathbf{V}^0$, $\mathbf{d}^0$, $\mathbf{X}$}
    \While{not converged}
        \State \textit{/* Update factor representation */} 
        \State $\mathbf{F}^{k+1} \gets$ 
        \text{Gradient step on factor dynamics}

        \State $\mathbf{B}^{k+1} \gets$ 
        \text{Iterative Sylvester solve for exposure profiles~\cite{bartels_algorithm_1972}}

        \State $\mathbf{Q}^{k+1} \gets$ 
        \text{Projection on Stiefel manifold~\cite{absil_matrix_2008}}

        \Statex
        \State \textit{/* Update graph representation, using the updated exposures $\mathbf{B}^{k+1}$ */}

        \State $\mathbf{W}^{k+1} \gets$
        \text{Sherman--Morrison solve for graph topology~\cite{golub_matrix_2013}}
    
        \State $\mathbf{d}^{k+1} \gets$ 
        \text{Quadratic-root solve of optimality condition}

        \State $\mathbf{V}^{k+1} \gets$ 
        \text{Proximal step enforcing sparsity}

        \Statex
        \State \textit{/* Update dual variables via gradient ascent */} 
        \State $\boldsymbol{\Lambda}_\mathbf{B}^{k+1} \gets \boldsymbol{\Lambda}_\mathbf{B}^k + \rho_\mathbf{B} (\mathbf{B}^{k+1} - \mathbf{Q}^{k+1})$
        \State $\boldsymbol{\Lambda}_\mathbf{W}^{k+1} \gets \boldsymbol{\Lambda}_\mathbf{W}^k + \rho_\mathbf{W} (\mathbf{W}^{k+1} - \mathbf{V}^{k+1})$
        \State $\boldsymbol{\Lambda}_\mathbf{d}^{k+1} \gets \boldsymbol{\Lambda}_\mathbf{d}^k + \rho_\mathbf{d} (\mathbf{W}^{k+1}\mathbf{1} - \mathbf{d}^{k+1})$

        \Statex
        \State $k \gets k + 1$
    \EndWhile
    \end{algorithmic}
\end{algorithm}

Direct optimisation of \eqref{eq:method_unified_obj} is challenging, as the smoothness term couples $\mathbf{B}$ and $\mathbf{W}$ and prevents them from being optimised independently. This motivates alternating schemes that separately navigate the different components of the objective, namely the differentiable terms, the proximal $\ell_1$ step, and the graph topology constraints. For this setting, the \acf{ADMM} is particularly appropriate~\cite{boyd_distributed_2010,cardoso_learning_2020}, as it introduces \emph{auxiliary variables} and decouples the optimisation into smaller tractable subproblems. Within this \acs{ADMM} framework, our unified \frameworkShort\ optimisation problem, defined in~\eqref{eq:method_unified_obj}, becomes
\begin{equation}
    \begin{aligned}
        \min_{\substack{
            \mathbf{B},\,\mathbf{F},\,\mathbf{W},\\
            \mathbf{Q},\,\mathbf{V},\,\mathbf{d}
        }} \quad
            & \lVert \mathbf{X} - \mathbf{B}\mathbf{F} \rVert_{\boldsymbol{\Omega}}^2
                + \lambda \sum_{i=1}^p \sum_{j=1}^p
                w_{ij}\,\lVert \mathbf{b}_{i} - \mathbf{b}_{j} \rVert_2^2 \\
            & + \alpha \lVert \mathbf{V} \rVert_1
                - \beta \; \mathbf{1}^\top \log \left(\mathbf{d} + \varepsilon \mathbf{1} \right) \\[0.3em]
            & + \delta \Vert \offdiag(\mathbf{F} \boldsymbol\Omega \mathbf{F}^\top) \Vert_F^2 \\[0.4em]
        \text{s.t.} \quad
            & \mathbf{Q}^\top \mathbf{Q} = \mathbf{I},\; \forall i, j \; v_{ij} \geq 0, \; \forall i\; v_{ii} = 0, \; \mathbf{V}^\top = \mathbf{V}, \\[0.2em]
            &\mathbf{Q} = \mathbf{B}, \; \mathbf{W} = \mathbf{V}, \; \mathbf{d} = \mathbf{W1},
    \end{aligned} 
    \label{eq:method_obj_aux}
\end{equation}
where \mbox{$\mathbf{Q} \in \mathbb{R}^{p \times k}$} is an orthogonality-constrained copy of $\mathbf{B}$, \mbox{$\mathbf{V} \in \mathbb{R}^{p \times p}$} is a sparse topologically constrained copy of $\mathbf{W}$, and \mbox{$\mathbf{d} \in \mathbb{R}^p$} is an explicit degree variable for the vector of node degrees, $\mathbf{W1}$. The augmented Lagrangian of \eqref{eq:method_obj_aux} can then be minimised by alternating over factor, graph, and dual-variable blocks. Within each iteration, the factor block is updated first, such that the graph topology adapts to the most recently estimated exposures. The ascent on the three dual variables, $\boldsymbol{\boldsymbol{\Lambda}}_\mathbf{B}$, $\boldsymbol{\boldsymbol{\Lambda}}_\mathbf{W}$ and $\boldsymbol{\boldsymbol{\Lambda}}_\mathbf{d}$, accumulates a history of primal--auxiliary misalignment and drives the variables towards consensus. In the iterative sequence described in~\autoref{algo:method_steps_admm}, $\rho_\mathbf{B}$, $\rho_\mathbf{W}$, and $\rho_\mathbf{d}$ denote the penalty parameters associated with the consensus constraints, \mbox{$\mathbf{Q} = \mathbf{B}$}, \mbox{$\mathbf{V} = \mathbf{W}$}, and \mbox{$\mathbf{d} = \mathbf{W1}$}. \\

\remark{The unified objective in~\eqref{eq:method_unified_obj} is non-convex, due to the bilinear reconstruction term $\lVert \mathbf{X} - \mathbf{B}\mathbf{F} \rVert_{\boldsymbol{\Omega}}^2$. As such, the \ac{ADMM} is not guaranteed to converge to a global minimum.}

\section{Implementation} \label{sec:implementation}
\textbf{Dataset.} For each of the S\&P 500, Nikkei 225, and STOXX 600 indices, the 100 most actively traded constituents of 2017 were selected. This yielded an asset universe of 300 equities, whose daily close prices were extracted from the Bloomberg Terminal in USD. \\

\noindent \textbf{Data Preprocessing.} The factor--graph representation was trained on log-returns, to ensure additivity of relative prices. Simple returns, which reflect profit and loss, were used for the calculation of in-sample and out-of-sample portfolio performance. \\

\noindent\textbf{\frameworkShort\ configuration.} The primal variables, $\mathbf{B}^0$ and $\mathbf{F}^0$, were initialised as the best rank-$k$ approximation of the downweighted historical returns. Specifically, this \acs{SVD}, in the form
\begin{equation}
    \mathbf{X}\boldsymbol{\Omega}^{\frac{1}{2}} = \mathbf{U} \boldsymbol{\Sigma} 
    \mathbf{V}^\top,
\end{equation}
yields $\mathbf{B}^0 = \mathbf{U}_{:, 1:k}$ and $\mathbf{F}^0 = \boldsymbol{\Sigma}_{1:k, 1:k} \mathbf{V}_{:,1:k}^\top \boldsymbol{\Omega}^{-\frac{1}{2}}$. The initial graph topology, $\mathbf{W}^0$, was constructed by applying a \ac{RBF} kernel to the initial exposures, $\mathbf{B}^0$, to give
\begin{equation}
    w_{ij}^0 = \exp \left( - \frac{\Vert \mathbf{b}_i^0 - \mathbf{b}_j^0 \Vert_2^2}
    {2 \hat{\sigma}^2} \right), \quad w_{ii}^0 = 0,
\end{equation}
where $\hat{\sigma}^2$ is the median pairwise squared distance between exposure profiles.
The auxiliary variables were initialised at their corresponding primal values, $\mathbf{Q}^0 = \mathbf{B}^0$, $\mathbf{V}^0 = \mathbf{W}^0$, $\mathbf{d}^0 = \mathbf{W}^0\mathbf{1}$. The dual variables, $\boldsymbol{\Lambda}_\mathbf{B}^0$, $\boldsymbol{\Lambda}_\mathbf{W}^0$, and $\boldsymbol{\Lambda}_\mathbf{d}^0$, were set to zero to reflect this initial primal--auxiliary consensus. \\

\noindent \textbf{Allocation Configuration.} The learnt representation can be assessed through portfolio allocation performance. To this end, we applied the Adaptive \acs{CutV} algorithm~\cite{dionysiou_adaptive_2026} to three configurations that incrementally incorporated components of the \frameworkShort\ representation. The first configuration served as a baseline, and performed allocation on the sample covariance matrix, $\SampleCovar$; the second replaced the sample covariance with the representation-induced covariance, $\RepCovar$, defined as in~\eqref{eq:bg_cov_apt}; the third leveraged both the induced-covariance and the graph component. In this final configuration, we considered a contagion-based perspective to guide the graph incorporation: if assets belonging to the same cluster share similar risk exposures, then the allocation should favour \emph{peripheral assets} which are comparatively less exposed to risk than the central ones~\cite{pozzi_spread_2013,peralta_network_2016}. As such, rather than allocating capital equally within leaves, we weighted assets by \emph{within-cluster weighted degree}, formalised by
\begin{equation}
   p_i \propto \frac{1}{\sum_{j \in \mathcal{C}} {w_{ij}}}
\end{equation}
where $w_{ij}$ denotes the edge weight connecting asset $i$ to asset $j$, both belonging to the leaf $\mathcal{C}$.
We refer to this allocation strategy as the \emph{Contagion Cut}. \\

\noindent\textbf{Backtest Configuration.} Rebalancing followed a rolling scheme, whereby the graph--factor representation was re-estimated monthly from a two-year lookback window, and was evaluated over the subsequent one-month period. \\

\autoref{tab:implementation_hyperparameters} summarises the objective and configuration hyperparameters. The performances reported in \autoref{sec:results} span the held-out period of \mbox{January 2019--March 2026}.

\begin{table}[h]
\centering
\caption{Hyperparameters of the proposed joint factor–graph framework. Parameters $k$, $\alpha$, and $C$ were selected using portfolio-level metrics, and $\lambda$, $\delta$, and $T$ using representation-level metrics; $\omega$ was fixed for a one-year half-life, and $\beta$ was fixed so that tuning $\alpha$ controls the sparsity–degree trade-off.}
\label{tab:implementation_hyperparameters}
\footnotesize
\begin{tabular}{llll}
\toprule
\textbf{Hyperparameter} & \textbf{Value} & \textbf{Role}\\
\midrule
Number of factors $k$ & $6$ & 
    Controls factor model's expressiveness\\[2pt]
Sparsity penalty $\alpha$ & $2.2$ & 
    Penalises over-connectivity  \\[2pt]
Recursive cuts count $C$ & $24$ & \acs{CutV} number of clusters  \\
\midrule
Smoothness strength $\lambda$ & $0.1$ & 
    Couples the graph and factor models \\[2pt]
Decorrelation strength $\delta$ & $1000$ & 
    Penalises redundancy in factors\\[2pt]
Lookback window $T$ & $2$ years & 
    Number of observations for training \\
\midrule
Degree barrier $\beta$ & $3.0$ & 
    Prevents degenerate, isolated nodes\\[2pt]
Temporal decay factor $\omega$ & $0.997$ & 
    Downweights older observations \\[2pt]
 Rebalancing frequency & Monthly & 
    Representation re-estimation rate \\
\bottomrule
\end{tabular}
\end{table}

\section{Experimental Results} \label{sec:results}
The learnt representation was evaluated as an economically meaningful, temporally persistent description of the market structure, and through its downstream utility in portfolio allocation. \\

\remark{For the regime breakdown, each trading day was assigned a market regime according to the daily level of the CBOE \acs{VIX}~\cite{noauthor_cboe_2026}: \emph{Calm} below 20, \emph{Elevated} between 20 and 30, and \emph{Crisis} above 30.}

 % STABILITY IMAGE MOVED TO IMPLEMENTATION, FOR IT TO SHOW AT TOP OF PAGE.
\begin{figure*}[htb]
    \centering
    \includegraphics[width=\textwidth]{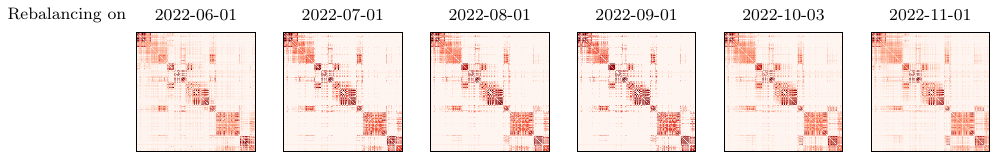}
    \Description{Stability (temporal NMI 0.768)}
    \caption{Learnt weight matrices of the exposure similarity graph across six consecutive rebalancing windows, with assets held in a fixed reference ordering to reflect genuine topological changes. The block diagonal structure persists across rebalancings, with a temporal \ac{NMI} of 0.768 over the full backtest period.}
    \label{fig:results_stability}
\end{figure*}

\subsection{Representation quality} \label{sec:eval_rep}

\begin{figure*}[tbp]
    \centering
    \begin{minipage}[t]{0.485\textwidth}
        \vspace{0pt}
        \includegraphics[width=0.984\linewidth]{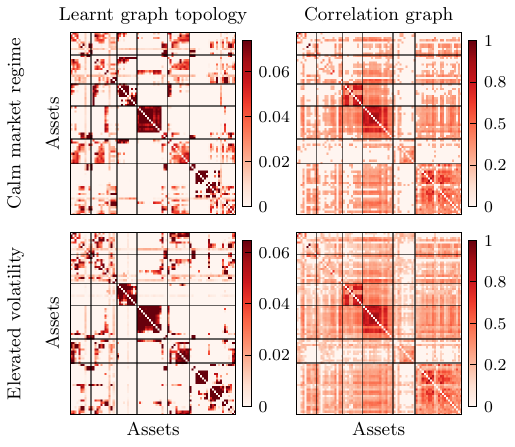}
        \caption{Weight matrices of the learnt exposure similarity graph, $\mathbf{W}_B$, and the correlation similarity graph, $\vert \SampleCovar \vert$, for the S\&P500, reordered by \acs{GICS} membership. The learnt topology aligned better with established economic taxonomy, with the improved alignment persisting across all CBOE \acs{VIX}-defined market regimes. Specifically, the within-cluster edge weights were $5.30\times$ stronger than between-cluster ones for $\mathbf{W}_B$ versus $2.42 \times$ for $\vert \SampleCovar \vert$ in \emph{Calm} markets, and $4.49 \times$ versus $1.64\times$ for \emph{Elevated} volatility regimes.}
        \label{fig:results_adjacency}
        \Description{}
    \end{minipage}\hfill
    \begin{minipage}[t]{0.485\textwidth}
        \vspace{0pt}
        \includegraphics[width=\linewidth]{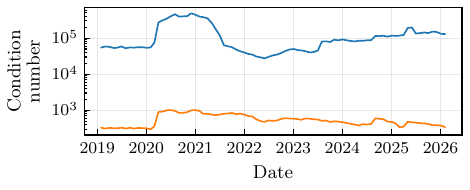}
        \includegraphics[width=\linewidth]{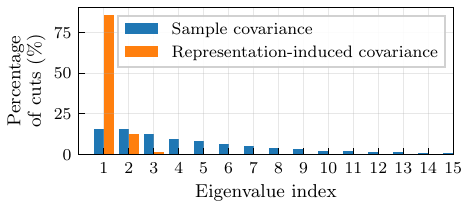}
        \caption{Conditioning of the representation-induced covariance, $\RepCovar$, relative to the sample covariance, $\SampleCovar$, and its consequence for partition recovery. \emph{Top:} Condition number over the backtest period. The representation-induced covariance remained roughly two orders of magnitude better-conditioned  than $\SampleCovar$. \emph{Bottom:} Distribution of eigenvalue indices selected by adaptive spectral cut. The better conditioning of $\RepCovar$ induced Adaptive \acs{CutV} to select the Fiedler vector (index 1) in 85.9\% of windows, $5.4 \times$ more frequently than with $\SampleCovar$.}
        \label{fig:results_fielder}
        \Description{}
    \end{minipage}
\end{figure*}

\noindent\textbf{Temporal stability.} Clustering on the learnt graph representation yielded persistent block diagonal structure over time. This is visualised in~\autoref{fig:results_stability}, where communities tend not to reshuffle across successive rebalancings. These stable community assignments are expected to reduce unnecessary turnover in~\autoref{sec:eval_port}. \\

% STABILITY IMAGE MOVED TO IMPLEMENTATION, FOR IT TO SHOW AT TOP OF PAGE.
\noindent \textbf{Interpretability.} The proposed framework learnt an economically meaningful structure, as it recovered the \acf{GICS} sectors~\cite{msci_global_2023} while remaining agnostic to the partitioning during training. This is illustrated in~\autoref{fig:results_adjacency}. \\

% IDEA 3: Denoising
\noindent\textbf{Covariance denoising.} Imposing a low-rank prior on the representation induced covariance, $\RepCovar$, reduces the number of parameters required for estimation, and thus reduces the estimation error in finite-data settings. This lower sensitivity to artefacts improves the spectral representation, as the eigenvalues of $\RepCovar$ are less compressed within the Marchenko--Pastur bulk than those of $\SampleCovar$. This is reflected in~\autoref{fig:results_fielder}, where $\RepCovar$ exhibited a condition number two orders of magnitude lower than that of $\SampleCovar$. Moreover, Adaptive \acs{CutV} recovered the optimal partition~\cite{shi_normalized_2000} $5.4\times$ more frequently with $\RepCovar$ than with $\SampleCovar$. Taken together, these findings suggest that the factor--graph representation effectively \emph{denoises} the covariance estimate, with such denoised estimator improving the performance of downstream allocation in~\autoref{sec:eval_port}.

\begin{figure*}[htb]
    \centering
    \includegraphics[width=\textwidth]{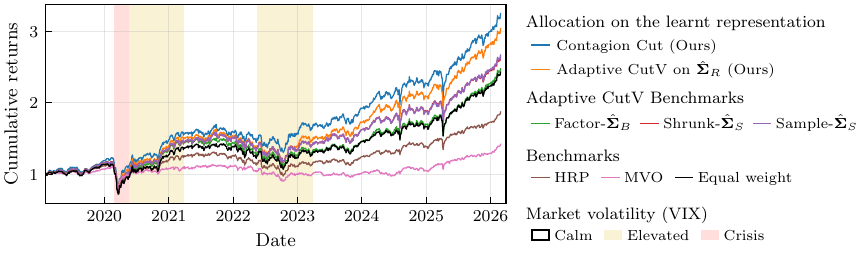}
    \caption{Cumulative returns of allocations on the learnt representation (Contagion Cut, Adaptive \acs{CutV} on $\RepCovar$) against the benchmarks (Adaptive \acs{CutV} on $\FactorCovar$, shrunk~$\SampleCovar$, and $\SampleCovar$; Equal weight), over the Jan 2019--March 2026 backtest period. The proposed Contagion Cut consistently outperforms its covariance-based counterparts, followed by Adaptive \acs{CutV} on $\RepCovar$.}
    \label{fig:results_cumulative_returns}
    \Description{}
\end{figure*}

\begin{table*}[ht]
\centering
\caption{Portfolio performance under transaction costs applied to one-way portfolio turnover, with CAGR in \%. The best value is shown in \textbf{bold}, while the second-best is shown \underline{underlined}. CutV is evaluated on the sample covariance, $\SampleCovar$, its Ledoit--Wolf (LW) shrunk counterpart~\cite{ledoit_well-conditioned_2004}, the factor-induced covariance, $\FactorCovar$, and the representation-induced covariance, $\RepCovar$. At all cost levels, performance increased monotonically from Adaptive CutV on $\SampleCovar$ to $\RepCovar$ to \mbox{Contagion Cut}.}
\label{tab:perf_tc}
\footnotesize
\setlength{\tabcolsep}{2.6pt}
\begin{tabular}{lrrrrrrrrrrrrrrr}
\toprule
 & \multicolumn{3}{c}{0\,bps} & \multicolumn{3}{c}{5\,bps} & \multicolumn{3}{c}{10\,bps} & \multicolumn{3}{c}{20\,bps} & \multicolumn{3}{c}{50\,bps} \\
\cmidrule(lr){2-4} \cmidrule(lr){5-7} \cmidrule(lr){8-10} \cmidrule(lr){11-13} \cmidrule(lr){14-16}
\textbf{Strategy} & CAGR & Sharpe & Calmar & CAGR & Sharpe & Calmar & CAGR & Sharpe & Calmar & CAGR & Sharpe & Calmar & CAGR & Sharpe & Calmar \\
\midrule
Equal weight & 15.5 & 0.780 & 0.418 & -- & -- & -- & -- & -- & -- & -- & -- & -- & -- & -- & -- \\
\midrule
MVO & 5.86 & 0.364 & 0.210 & 5.77 & 0.356 & 0.207 & 5.67 & 0.348 & 0.203 & 5.48 & 0.333 & 0.196 & 4.92 & 0.287 & 0.176 \\
MVO, LW-$\SampleCovar$ & 6.21 & 0.390 & 0.222 & 6.13 & 0.383 & 0.219 & 6.05 & 0.376 & 0.216 & 5.88 & 0.362 & 0.210 & 5.38 & 0.321 & 0.192 \\
HRP & 10.7 & 0.613 & 0.324 & 10.6 & 0.606 & 0.321 & 10.5 & 0.600 & 0.318 & 10.3 & 0.586 & 0.311 & 9.67 & 0.545 & 0.292 \\
\midrule
CutV, $\SampleCovar$ & 17.3 & 0.860 & 0.480 & 17.1 & 0.854 & 0.476 & 17.0 & 0.848 & 0.473 & 16.8 & 0.836 & 0.466 & 16.1 & 0.800 & 0.445 \\
CutV, LW-$\SampleCovar$ & 17.1 & 0.856 & 0.474 & 17.0 & 0.850 & 0.470 & 16.9 & 0.843 & 0.467 & 16.6 & 0.831 & 0.460 & 15.9 & 0.794 & 0.440 \\
CutV, $\FactorCovar$ & 15.8 & 0.788 & 0.433 & 15.7 & 0.783 & 0.430 & 15.6 & 0.778 & 0.427 & 15.4 & 0.768 & 0.422 & 14.8 & 0.738 & 0.405 \\
\midrule
\textbf{CutV, $\RepCovar$ (Ours)} & \underline{19.7} & \underline{0.994} & \underline{0.580} & \underline{19.5} & \underline{0.983} & \underline{0.573} & \underline{19.3} & \underline{0.972} & \underline{0.566} & \underline{18.8} & \underline{0.949} & \underline{0.552} & \underline{17.5} & \underline{0.883} & \underline{0.510} \\
\textbf{Contagion Cut (Ours)} & \textbf{21.0} & \textbf{1.07} & \textbf{0.611} & \textbf{20.8} & \textbf{1.05} & \textbf{0.603} & \textbf{20.5} & \textbf{1.04} & \textbf{0.594} & \textbf{19.9} & \textbf{1.01} & \textbf{0.578} & \textbf{18.3} & \textbf{0.929} & \textbf{0.528} \\
\bottomrule
\end{tabular}
\end{table*}

\subsection{Portfolio performance} \label{sec:eval_port}

\noindent\textbf{Representation role in performance.} \autoref{tab:perf_tc} gives the isolated contribution of each component of the joint factor--graph model. The Adaptive \acs{CutV} applied to the representation-induced covariance, $\RepCovar$, consistently outperformed the sample, $\SampleCovar$, and factor-induced, $\FactorCovar$, alternatives. This performance gain is consistent with two previous observations: the improved conditioning of $\RepCovar$ favoured cuts at the optimal partitioning, while the objective of~\eqref{eq:method_unified_obj} regularised traditional \acs{PCA}--factor approaches to capture structure beyond variance maximisation.
The further gains from Contagion Cut show that the graph topology contributed to the portfolio performance on top of the better-conditioned covariance estimate. The monotonic improvement in performance is also shown in~\autoref{fig:results_cumulative_returns}. \\

\remark{In~\autoref{tab:perf_tc} and~\autoref{fig:results_cumulative_returns}, \ac{MVO} and \ac{HRP} underperformed, as they respectively amplified the estimation noise in $\SampleCovar$ and relied on a locality measure that conflates systematic and idiosyncratic risk. Equal weights outperformed both, but remained agnostic to historical returns. Portfolios derived from the learnt representation consistently outperformed the benchmarks, reaffirming the benefit of a market-aware allocation.} \\

\begin{figure}[htb]
    \centering
    \includegraphics[width=\columnwidth]{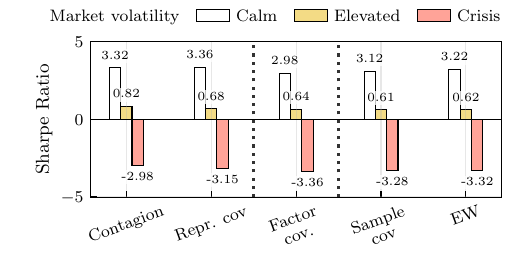}
    \caption{Annualised Sharpe Ratio of each portfolio within each \acs{VIX}-derived volatility regime, computed over all trading days assigned to that regime.
    The representation-based portfolios outperformed the benchmarks in \emph{Calm} and \emph{Elevated} volatility regimes, with Contagion Cut registering the least negative Sharpe ratio in \emph{Crisis}.}
    \label{fig:results_regime_perf}
    \Description{}
\end{figure}

\noindent\textbf{Regime-conditional performance.}  \autoref{fig:results_regime_perf} shows the decomposition of portfolio performance by \acs{VIX}-classified market regime. Contagion Cut and Adaptive \acs{CutV} on $\RepCovar$ outperformed the baselines in \emph{Calm} and \emph{Elevated} markets. In \emph{Crisis}, representation-based approaches registered the least negative Sharpe ratio. This negative Sharpe reflects the long-only constraint, which cannot profit from a market-wide decline.\\

\begin{figure*}[htb]
    \centering
    \includegraphics[width=\textwidth]{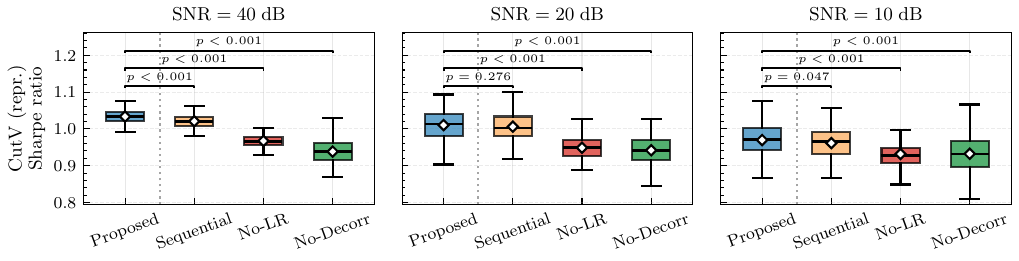}
    \caption{Effect of each term of the \frameworkShort\ objective in~\eqref{eq:method_unified_obj} on downstream allocation performance, with a t-paired testing over $100$ seeds. Performance is quantified through the Sharpe ratio of adaptive \acs{CutV} on the resulting induced covariance. Low, medium and high noise corruption settings are considered with a \ac{SNR} of approximately 40, 20, and 10 dB. Removing the low-rank prior (No-LR~\cite{dong_learning_2016}) or the non-redundancy constraint (No-Decorr) significantly degrades performance at every noise level ($p < 0.001$). Sequential estimation of the factor then graph domain significantly degrades performance for the low noise, with statistical significance vanishing for moderate and high noise levels.}
    \label{fig:results_ablation_significance}
    \Description{}
\end{figure*}

\remark{The learnt representation models historical returns as arising from $k$ factors. In \emph{Crisis}, returns load onto fewer risk drivers~\cite{longin_extreme_2001}, yet the model is constrained to populate all $k$ dimensions. This can introduce artefacts into the learnt representation, rendering the factor model least reliable when diversification is most needed. As such, the proposed framework does not fully mitigate \emph{Crisis} risk; instead, its primary outperformance is concentrated in \emph{Calm} and \emph{Elevated} volatility regimes, where capital is predominantly compounded.} \\

\noindent\textbf{Effect of each regularisation term.} We studied the effect of each regularisation term of~\eqref{eq:method_unified_obj} on downstream allocation performance, and evaluated whether this holds under varying levels of estimation noise. To this end, synthetic Gaussian noise of variance $\sigma^2$ was added directly to the raw returns, $\mathbf{X}$, before estimating the factor--graph model. The results are shown in~\autoref{fig:results_ablation_significance}. The low-rank prior and the decorrelation term are both necessary, as performance significantly degraded in their absence ($p < 0.001$). For the low noise setting, coupling the factor and graph domains significantly outperformed their sequential estimation, indicating that the domains became cross-aware and successfully mutually regularised one another towards an improved, shared market view. This performance gain, however, lost its statistical significance under high noise corruption, where mutual consistency alone might not suffice against prevailing sampling noise.

% \section{Discussion} \label{sec:discussion}
% \input{5_discussion}

\section{Conclusion}
Market representations underpin portfolio diversification, yet existing approaches typically rely on either factor models, which capture systematic sources of risk, or graph-based models, which capture inter-asset relationships. Our results have shown that these complementary market views can be reconciled to provide a stronger basis for diversification. Specifically, we introduced the \frameworkLong, which jointly learns the factor and graph domains by redefining locality through systematic exposure profiles, rather than via observed co-movements. 
This ensures that neither domain is estimated in isolation. 
Although the resulting optimisation problem is non-convex, an \ac{ADMM} formulation has yielded a tractable and scalable solution. The resulting representation captures stable systematic relationships, rather than transient pairwise co-movements, producing graph structures that are temporally persistent and consistent with known economic taxonomy.

These results have demonstrated that systematic risk and inter-asset relationships should not be treated as competing views of the market, but rather as complementary domains. The joint learning paradigm has introduced a cross-domain awareness that mutually regularises both the factor and graph domains. This thus yields a richer market representation which, in turn, drives an improvement in portfolio performance across regimes and transaction costs. The proposed exposure-based locality therefore offers a more informative notion of market structure than correlation-based similarity, and consequently constitutes a stronger basis for portfolio diversification.
More broadly, these findings open avenues towards mutually-aware factor–graph models, wherein complementary representations inform and regularise one another.

% \begin{acks}
%     \input{acknowledgments}
% \end{acks}

% \section*{Ethics and Privacy Statement}
% \input{ethics}

\bibliographystyle{ACM-Reference-Format}
\bibliography{references_frozen}

\end{document}